\documentclass[useAMS,usegraphicx]{mn2e}

\usepackage{amsmath,dcolumn,threeparttable,booktabs}
\usepackage{color}

\title[Assessing the complexity of orbital
parameters after asymmetric kick in binary pulsars] {Assessing the complexity of orbital
parameters after asymmetric kick in binary pulsars} \vskip 3cm

\author[Taani et al. ] {Ali Taani$^{1}$\thanks {E-mail: ali.taani@bau.edu.jo}, Juan C. Vallejo$^{2,3}$ and Mohammed Abu-Saleem$^{4}$\\
$^{1}$Physics Department, Faculty of Science, Al-Balqa Applied University, 19117 Salt, Jordan \\
$^{2}$Joint Center for Ultraviolet Astronomy, AEGORA Research Group, Universidad Complutense de Madrid, Avda Puerta de Hierro,  Madrid, Spain\\
$^{3}$Nonlinear Dynamics, Chaos and Complex Systems Group, Departamento de F\'{i}sica, Universidad Rey Juan Carlos, Tulip\'{a}n,  M\'{o}stoles, Spain\\
$^{4}$Mathematics Department, Faculty of Science, Al-Balqa Applied University, 19117 Salt, Jordan\\}

\begin{document}
\date{Received date ; accepted date}

\maketitle
\begin{abstract}

The dynamical characterization of the Millisecond Pulsar (MSP) parameters is a key issue in understanding these systems.
We present an analytical analysis of the orbital parameters of binary MSPs
with long periods (P$\rm_{orb} > 2$ d) and circular ($e \leq $ 0.1) orbits, produced by an asymmetric kick model imparted during the Accretion Induced Collapse (AIC) 
of white dwarfs process.
It turns out that the distribution of orbits peaks up to P$_{orb,f} \leq $ 90 d with  strong circularization. Considering the different assumptions 
about the distribution of companion He stars $ 3M_{\odot}\leq M_{com} \leq5M_{\odot}$, the binary will affect the setups of the balance condition of minimum energy.

Our analytical approach is just a first approach to the more complete models required for describing all binary parameters after an asymmetric kick.
Therefore, we have also run some numerical simulations
in order to compare their results  with the
analytical studies. We aim to initiate a first
exploration of the full complexity of the problem,
when combining a variable kick time and a variable kick vector direction.
Indeed, the
numerical simulations show patterns resembling the complex behavior
found in chaotic scattering problems. Although we deal with a deterministic problem and bounded orbits, the regular characteristic orbits are found in more realistic phases during the AIC process. In addition, the overall process can show complex behaviors
strongly associated with the internal kick mechanisms. This would lead us to identify the nature of regular orbits and their orbital morphology.

\end{abstract}



\section{Introduction}

The discovery of Millisecond Pulsars (MSPs) has led to fresh
insights into the astrophysics of Neutron Stars (NSs) (see e.g. Bhattacharya \& van den Heuvel 1991; Tauris \& van den Heuvel 2009; Lorimer 2009; Taani et al. 2012a,b; Tauris 2015; Taani \&  Khasawaneh 2017).
 Interestingly, the Accretion Induced Collapse (AIC) of massive white dwarfs  (Nomoto 1987; Isern \& Hernanz 1994; Ferrario \& Wickramasinghe 2007; Wickramasinghe et al. 2009; Hurley et al. 2010; Zhang et al. 2011; Chen et al. 2011; Taani et al.
2012b, Tauris et al. 2013; Freire \& Tauris 2014; Ablimit \& Li 2015a; Tauris 2015; Kwiatkowski 2015; Tauris et al. 2017; Ablimit 2021a,b) has been invoked to explain the formation of some binary MSPs, instead of the standard model (Alpar et al. 1982; Zhang \& Kojima 2006; Taani et al. 2018; Taani et al. 2019).

It has been studied that the magnetized WDs can grow in mass with low accretion rates (Ablimit \& Maeda 2019). Ablimit (2021b) showed that the channel of magnetized ONe or CO) WD + stripped helium star binaries can produces MSPs, magnetars and type Ia supernovae (SNe Ia which are thermonuclear explosions of CO WDs that have grown to the Chandrasekhar limit).

The magnitude and direction of asymmetric kicks imparted on NSs during AIC are one of the major uncertainties in stellar population synthesis studies. Note that there are other possible mechanisms that may affect the orbital evolution of the WD binaries such as tides, and post-AIC mass loss and accretion (e.g., Ablimit \& Li 2015; Tauris et al. 2013; Abu-Saleem  \& Taani 2021a,b; Wang et al. 2022). The physical
mechanisms that cause these kicks are still open questions,
but they are
presumably the result of some asymmetry in the  core collapse or
subsequent SNe explosion (see, e.g., Pfahl et al. 2002;
Podsiadlowski et al. 2004; Wei et al. 2010; Janka 2017; Tauris et al. 2017; Taani et al. 2019;  Taani et al. 2020).

On the observational side, many authors presented strong observational evidences like the optical infrared light curves and spectra of the nickel-rich outflows through the transient surveys (see i.e. Melatos 2007; Darbha et al. 2010; Piro \& Kulkarni 2012; Metzger \& Piro 2014; Piro \& Thompson 2014; Taani 2015; Metzger et al. 2009; Melatos et al. 2021; Gautam et al. 2021), many calculations of stellar models and their parametrization  in population  synthesis  codes can be found in Tutukov \& Yungelson (1993), Ablimit et al. (2019), Abdusalam et al.(2020), Ablimit (2021a) and many other similar binary population synthesis works.

We aim to provide some insight into the possible presence of complexity when asymmetric kicks processes are imparted on NSs during AIC.
We will focus on searching for the presence of dynamical chaos,
meaning strong dependency to the initial conditions of the system.
We will follow some of the ideas applied to chaotic scattering
studies. Indeed, in a broad sense, a scattering  problem
can be seen as the problem to obtain the relationship between some input variables, that characterize an initial condition for some dynamic system, and some output variables, that characterize the final state of the system.

Chaos can be present even in
very simple two-degrees of freedom deterministic dynamical systems, if one adds
a time-dependant process to the system.
In chaotic scattering problems, complex patterns
can be observed when the incoming particle can
find the target system in a state that depends on the arrival time.
That is, this state depends on the precise time at which the
incoming particle interacts with the target system.

Likewise these scattering processes, asymmetric kicks processes can add a similar time-dependency to exploding binaries.
The kick results will depend
on the precise state of the orbital parameters of the binary system
when the kick occurs.
Hence, the overall binary pulsars problem
can show strong sensitivity to initial
control parameters. These parameters can be, for instance,
the masses defining the binary system, or the
precise time when the explosion (kick) takes place.

Therefore, we have modified the regular two-body problem
that describes the WD-companion binary system, and we have included
a kicker process that will add the necessary time-dependency.
This kicker process will modify the orbital parameters at a given instant, and
it will allow the possibility of having a complex, may be chaotic, motion.

The structure of the paper is as follows. The Section 2
present the main binary characteristics and introduces
a simple analytical approach for the AIC process.
The Section 3 numerically explores the complex
behavior of these binaries when a variable kick process is introduced in a otherwise regular two-body system.
The Section 4 discusses those behaviors and compares the analytical
and numerical approaches. Finally, the last section summarizes the results and makes some concluding remarks.

\section{A simple analytical approach for the AIC process}

In the case of AIC, low kick velocities are expected, of the order of 50 km/s (Boyles et al. 2011). However,  van den Heuvel (2007) argued that (6/8) double NSs are known to have low eccentricities, suggesting that their second-born NSs received  very small kick velocities  at birth ($\leq 150 kms^{-1}$). In his work, he argued that these low-mass and low-kick NSs are most likely formed by the AIC. The change in binary parameters not only
occurs from mass loss and kick
but also increases in momentum from their formation mechanism 
(Arzoumanian et al. 2002; Kiel \& Hurley 2009). For an instant, changing in mass in AIC comes from converting baryonic mass to gravitational mass during the formation of the NS (see Zeldovich \& Novikov 1971; Bagchi 2011; Mardini et al. 2019a,b; Mardini et al. 2020; Almusleh et al. 2021). This physics is
described, among others, in Fryxell \& Arnett 1981 and Kalogera 1996.

As a very first approach to this problem, we present here an analytical
model to estimate the changes in orbital
periods and semi-major axes because of the kick velocity received by the WD when the AIC process
takes place. In our model, most of the mass that is lost in a binary system that undergoes AIC comes from the WD converting baryonic mass into binding energy during the collapse and the formation of the newborn NS, and the mass change is of the order of 0.18 M$_{\odot}$ (i.e.  Ablimit \& Li 2015).


The quantitative relationship between the orbital eccentricity and the amount of mass
ejected in the AIC is (see i.e., Eqs. 19-28 in Tauris \& Takens 1998),
\begin{equation}
e_{f} - e_{i}= \frac{\Delta M_{AIC}}{M_{final}}
\end{equation}
where $\Delta M_{AIC}$ is the mass ejected during the AIC.
Therefore, $\Delta M_{AIC}$ = M$_{ej}$ = M$_{WD}$ - M$_{NS}$. By other hand,
$M_{final}$ =  $M_{comp}$ + $M_{NS}$, the sum of the
masses of the companion and the NS.
In our calculations,
we assume that the binary orbits prior to the AIC process are circular, e$_{i}$ = 0.
To some extent, this is justified by the fact that
close binaries are very likely to circularize as a result of mass
transfer from the evolved star before the explosion (e.g. Tauris \& Takens 1998, Cai et al. 2012).  The companion star may lose material from its surface because the SN
ejecta may either directly strip material from the companion by
direct transfer of momentum or evaporate the envelope through the
conversion of the blast kinetic energy into internal heat (Li 2008).  Say M$_{rec}$ as the amount
of mass received by the companion ($\sim 0.1-0.2M_{\odot}$ see Zhang et al. 2011). Then
M$_{comp,f}$ = M$_{comp,i}$ + M$_{rec}$, where
M$_{comp,i}$ and M$_{comp,f}$ are the initial and final masses of the companion, respectively.

We can cancel the change in binding
energy here as this is not a NS.
Let us assume that M$_{rec} \ll M_{ej}$. As a first approach to the
problem we will analyse a planar $2D$ problem as seen in the Figure 1. In this case, we can deduce the relation of orbital period and new parameter $\eta$ (which is related with kick velocity) based on the previous assumptions.



\begin{figure}
\includegraphics[width = 8cm]{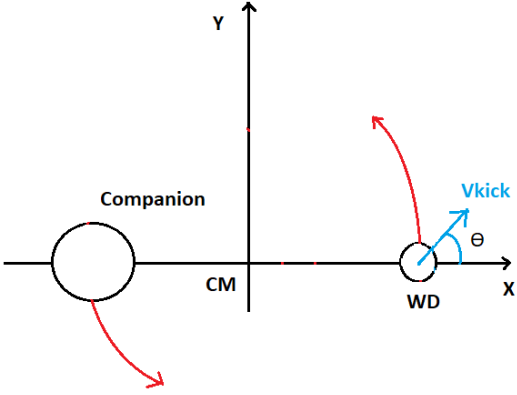}
\\
\caption{
A simple schema of the geometry of the planar binary system under
the effect of a supernova kick, the angular momentum vector
pointing along the positive Z-axis.
The WD explodes, which implies it receives a kick velocity vector
that forms a time-dependant angle $\theta$ with the X and Y axes.
A complex change of the initial orbital parameters
can occur when one varies the $t_{kick}$ time, or time
when the explosion takes place. However, it can also occur
when the initial masses of the binary are considered control parameters
and they are varied whilst $t_{kick}$ remains fixed.
Note that, once the kick take place,
the WD losses a given amount of mass $M_{ejec}$, while the companion receives
a different, smaller, amount of mass $M_{rec}$.}
\label{fig1}
\end{figure}



Before the AIC event the orbital energy can be expressed as (see, e.g. Hills 1983 and references therein).
\textbf{\begin{equation}
 E_{orb,i} =  \frac{- GM_{WD} M_{comp,i}}{2a_{i}}
 \end{equation}}

 where $G$ is the gravitational constant, a$_{i}$ is the initial separation between the two stars in the binary.




Hence, the orbital energy after the SNe,

\textbf{\begin{equation}
 E_{orb,f} = - \frac{G M_{NS} M_{comp,f}}{2a_{f}},
 \end{equation}}
with a$_{f}$ being the final separation.


By using the  ratio of initial to final velocities (the velocity of WD relative to it's companion before and after the AIC event (v$_{i}$/v$_{f}$)),  and the energy conservation law
in the equations above, we can get,

 \begin{equation}
 \frac{a_{f}}{a_{i}} = [\frac{1-\frac{\Delta M}{M_{b}}}{1-(2\frac{\Delta M}{M_{b}}) - \frac{v^{2}_{f} - v^{2}_{i}}{v^{2}_{orb}}}]
 \end{equation}

 where $\Delta M$= $\Delta M_{AIC} - \Delta M_{rec}$, $M_{b} = M_{WD}+M_{comp,i}$, and the orbital  velocity $v_{orb}=\sqrt{\frac{GM_{b}}{r}}$.


By some calculations we get,


 \begin{equation}
 \frac{v^{2}_{f} - v^{2}_{i}}{v^{2}_{orb}} = (\frac{v_{kick}}{v_{orb}})^{2} + 2(\frac{v_{i}}{v_{orb}}) (\frac{v_{kick}}{v_{orb}}) cos\theta
 \end{equation}

If we adopt $cos\theta = 1$, i.e. $v_{kick}$ (kick velocity) is co-aligned with the direction of the orbital angular momentum of the pre-explosion binary,

 \begin{equation}
 \frac{a_{f}}{a_{i}} = [\frac{1-\frac{\Delta M}{M_{b}}}{1-2(\frac{\Delta M}{M_{b}}) - (\frac{v_{kick}}{v_{orb}})^{2} - 2(\frac{v_{i}}{v_{orb}}) (\frac{v_{kick}}{v_{orb}})}]
 \end{equation}
 The orbital speed is defined by
   \begin{equation}
   v_{orb} = \sqrt{\frac{G(M_{WD}+M_{comp,i})}{r}} \sim \sqrt{\frac{GM_{WD}}{r}}
    \end{equation}





\begin{figure}
\includegraphics[width=8.0cm, angle=0] {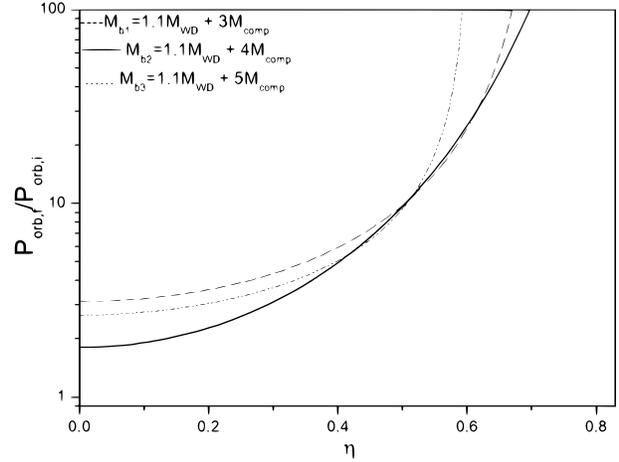}
\caption{The $\rm a_{f}/a_{i}$ as a function
of $\eta$. It is notable that when $\eta$ approaches 0.40, the
ratios increase sharply according to Eq. 8, (see the text). The
various assumptions of $\eta$ lead to different distances after the AIC process with different
lines.}
\label{eta1}
\end{figure}
\begin{figure}
\includegraphics[width=8cm, angle=0] {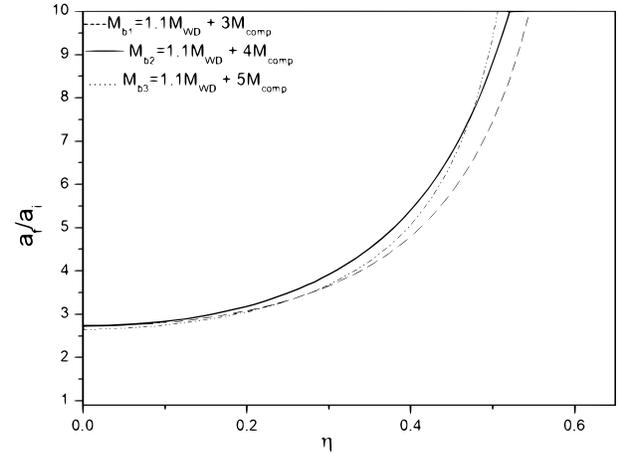}
\caption{ The same meaning as Fig. 2, but $\rm P_{orb,f}/P_{orb,i}$ as a function of $\eta$.} \label{eta2}
\end{figure}

We assume here that the corresponding $v_{orb} \sim v_{esc}$. The combination of kick components (i.e., the magnitude and direction) will be an effective means for determining their orbital separation, their orbital periods, and, of course, the fate of the stars (Tauris  \& van den Heuvel 2006; Taani 2016). This gives rise of $a_{f}$ to a critical value of $a_{f}$ = 3$a_{i}$ ($1-e_{i})$ at $e_{i} =0$. We assume that  $a_{i} = 1.2 R_{\odot}$, which might be reasonable for AIC binaries where the companion is on the MS of helium MS.

However, if the amount of angular momentum of the accreted material equals the orbital angular momentum, this would lead to the possibility of the companion being kicked away (Kiel et al. 2008). An assumption here is that a more
energetic explosion would be able to impart a greater initial
angular velocity onto the proto-NS than a less energetic explosion, and the
result is typically an eccentric orbit with an orbital period of a
few hours (Hurley et al. 2002).

Now let us define the $\eta$ parameter as the ratio between two velocities
\begin{eqnarray}
\label{b}
\eta=\frac{v_{kick}}{v_{esc}}
\end{eqnarray}

%


The ratio of final to initial orbital periods can be written as

 \begin{equation}
 \frac{P_{f}}{P_{i}} = [\frac{1-\frac{\Delta M}{M_{b}}}{1-2(\frac{\Delta M}{M_{b}}) - \eta^{2} - 2\eta}]^{\frac{3}{2}}
 \end{equation}

Figures~2 \& ~3 illustrate the behavior of these equations
as the $\eta$ value evolve by fixing
$\Delta M$  and the binary mass $M_b$.
We have taken an illustrative $\Delta M=M_{ej}-M_{rec}=0.18$, $M_{rec}$ is likely close to 0, and $M_{ej}$ is likely closer to 0.18$M_{\odot}$, that correspond to the mass of the WD that is converted into binding energy of the NS (see Zhang et al. 2011; Bagchi 2011). The binary mass was in turn computed using
a WD mass of $M_{WD} = 1.1 M_{\odot}$ and a companion mass of
$3M_{\odot}\leq M_{comp} \leq5M_{\odot}$.
Our sample scenario seems to be
reasonable for AIC binaries where
the companion is on the MS or helium MS
(see other possible scenarios in Ruiter et al. 2019).

\section{Numerical simulations and regular or chaotic processes
resulting from AIC}

In the previous sections, we have estimated the new orbital period
imparted to the newborn pulsar by relying on the kick velocity, considering an initial circular orbit.
However, this is a very special case. The general
problem can present a wider set of initial conditions,
and this section aims to show the complexity of the
calculations when considering a more complete case. An indicator of the presence of high complexity in a given  system is the detection of dynamical chaos,
or strong sensitivity to initial conditions (Alligood et al. 1996).
The presence of this phenomenon means
that even minor  differences in the values
of the initial conditions can lead to very different
final states of the system.

The standard two-body problem can be seen as one reduced-mass
particle. Hence, it can only show regular motions. This is because, following the Poincaré-Bendixon theorem, chaos cannot be present
in continuous flows of second order (Vallejo \& Sanjuan 2019).
However, if one introduces an additional dimension into the system,
the time in this case, complexity can arise.
In our case, this additional temporal dimension can be
introduced by means of the kick process.

For solving the numerical binary, we will use
a standard $N$-body solver, the Rebound package (Rein \&  Liu 2012).
This numerical engine integrates the motion of $N$-particles under the influence of gravity, allows the use of a variety of integrators,
and supports changes in orbital parameters at designated times.


Each simulation starts with given initial
conditions of an arbitrary, non-circular, binary system.
We have fixed the initial eccentricity to be
$e_0=0.01$, and the semi-major axis $a_0=1.0$au along all simulations.
We have also set $\Omega=0.0$, $\omega=0.0$ and $i=0$,
as we are dealing with a planar problem.

The masses of the two bodies will be
$M_{WD}$ for the exploding white dwarf, and $M_{comp}$ for the companion, with
$M_{comp} > M_{WD}$.
Once the explosion takes place,  at an arbitrary time $t_{kick}$,
the WD star losses $M_{ejec}$ mass and
suffers the necessary kick, changing its velocity vector.
The $M_{comp}$ mass corresponds to the initial mass of the companion star,
that gains some small amount of mass $M_{rec}$ after the explosion.

There is  firm evidence that many newborn NSs
receive this momentum kick at birth (Podsiadlowski et al.
2004), which can raise the observed high velocities (typically
$\sim$ 400 kms$^1$). Note that in AIC, low kick velocities are expected to be much lower than this value (see Boyles et al. 2011).

Chaotic behavior can be generated
by this kick, even for the lowest values of the kick.
This is because the system strongly depends on $t_{kick}$:
a kick produced when
the stars are in different orbital states will produce different final results.
Depending on the kick strength,
the resulting stars will or will not remain bounded.

Hence, the first control parameter to take into account
is the kick velocity vector $\vec{v}_{kick}$. This vector defines the recoil that the exploded star receives.
This will modify the orbital velocity of the kicked body,
that will have a final mass $M_{WD}-M_{ejec}$ after the explosion.

This kick vector is defined by its the modulus and its
direction, and the recoil velocity components
added to the kicked star will strongly depend on the
selection of this direction. The kick direction can depend
on many different factors on a real kick scenario.
As a first approach to it, we have defined the following
kick process. We have implemented a rotating vector,
with an arbitrary intrinsic period $T_{kicker}$.
At the precise kick time, $t_{kick}$, the
kick is modeled as a vector of modulus $v_{kick}$,
pointing to the direction of the rotating vector at that time.

Hence, depending on $t_{kick}$,
the kick velocity vector will be pointing to a different direction,
defined by a variable $\theta$ in Figure~1.
This kick will finally modify the orbit of the stars,
leading to different orbital parameters after the explosion.

The period of the rotating vector has been fixed to be $T_{kicker}=0.1$yrs
in all simulations, as we want to address
the complexity of the final dynamics and not the precise values
resulting from each simulation.
We will compare our numerical results with
the previous section analytical calculations. We will also analyse
the resulting complexity by focusing on the
values of the semi-major axis $a$ and the orbital period $T$.
In addition, we have also analysed the changes in eccentricity in some cases.

\begin{figure*}
\begin{center}
\begin{tabular}{cc}
\includegraphics[width=0.5\linewidth]{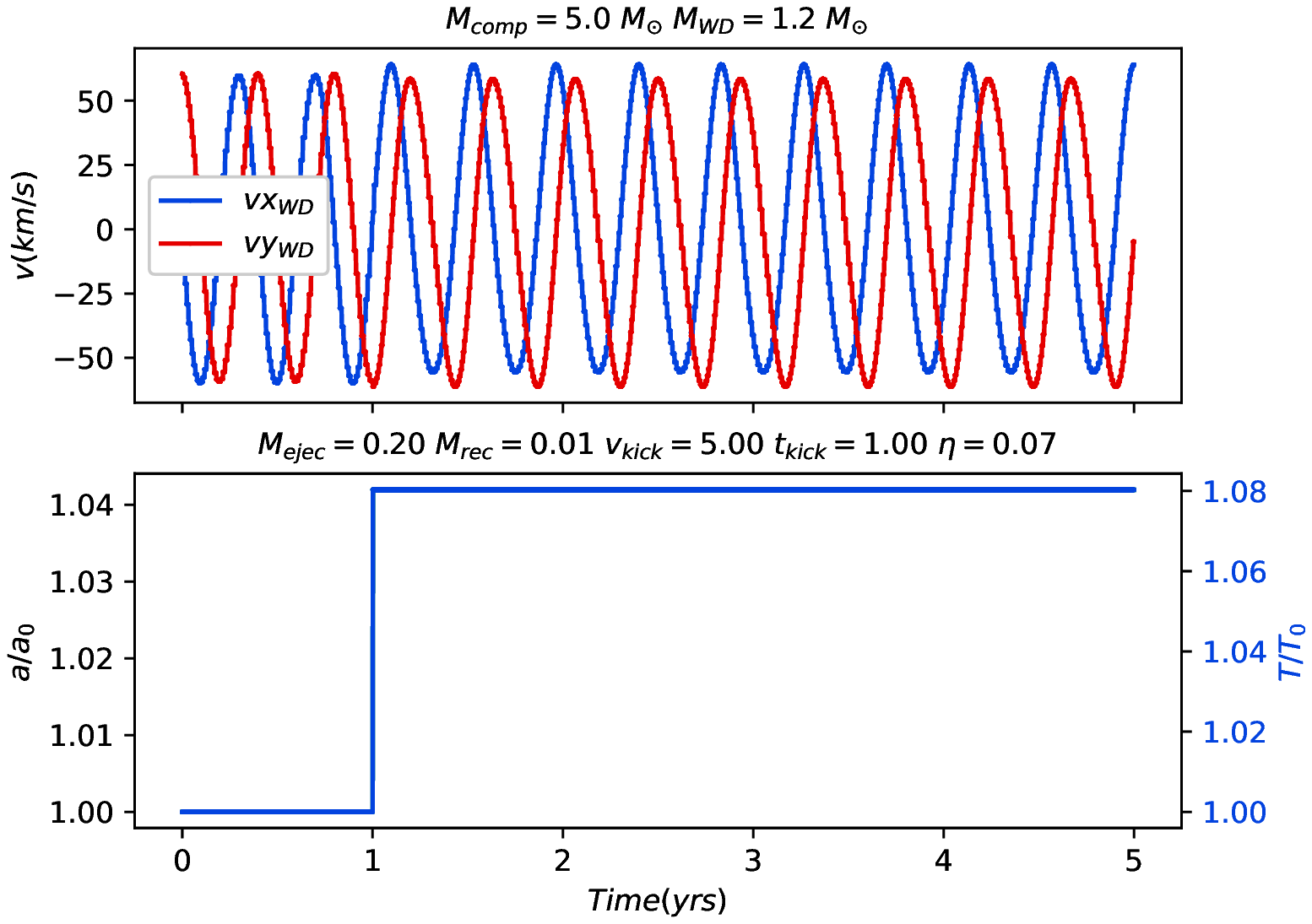} &
\includegraphics[width=0.5\linewidth]{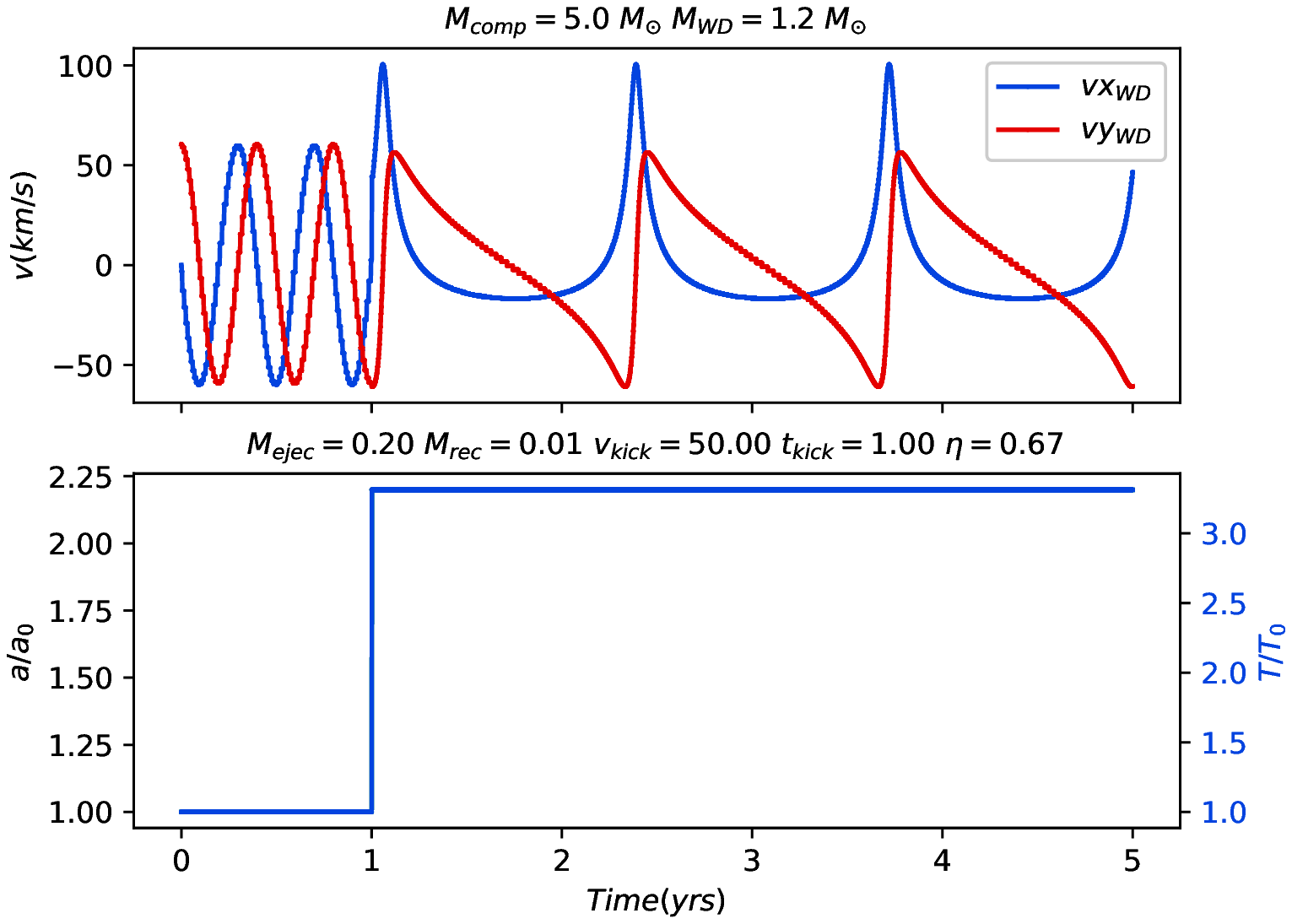}\\
\end{tabular}
\end{center}
\caption{The consequences of a kick
on a typical binary system (see system details in the text).
{The initial orbital period is $T_0=0.4$yrs.
The kick take place at $t_{kick}=1.0$yrs.
Upper diagrams show how the $v_x$ and $v_y$ components
for the white dwarf are modified after the kick.}
The bottom diagrams show the effect in semi-major axis
and orbital period.
Left panel corresponds to a kick velocity modulus
$v_{kick}=5.0$km/s, when the change is relatively small.
The right panel corresponds to $v_{kick}=50.0$km/s, when
the changes are larger. One can observe noticeable
changes in the final orbital parameters of the system in both cases.}
\label{kicker_examples}
\end{figure*}

The Figure~5
reflects a typical case of the impact of
a kick on the initial binaries.
The system masses are set
to be $M_{WD}=1.1$ $M_{\odot}$ and $M_{comp}=5.0$ $M_{\odot}$,
resulting on an initial period of $T_0=0.7$yrs.
The kick is defined
to happen at $t_{kick}=1$yr after the simulations starts.
Remaining parameters are set to be
$M_{ejec}=0.2 M_{\odot}$, $M_{rec}=0.01 M_{\odot}$.
The left panel corresponds to a kick velocity modulus
$v_{kick}=5.0$km/s, when the change is relatively small.
The right panel
corresponds to $v_{kick}=50.0$km/s, when
the changes are larger. One can observe noticeable
changes in the final orbital parameters of the system in both cases.
Figure~1 also shows that
the final period $T$ and semi-major axis $a$
increase, but the system remains bounded in both cases.
Hence, we have defined a fully deterministic dynamical system that can show a very complex behavior despite the simplicity of the
defined kick mechanism, and it even can present
strong dependency on initial conditions.

\begin{figure*}
\begin{center}
\begin{tabular}{cc}
\includegraphics[width=0.5\linewidth]{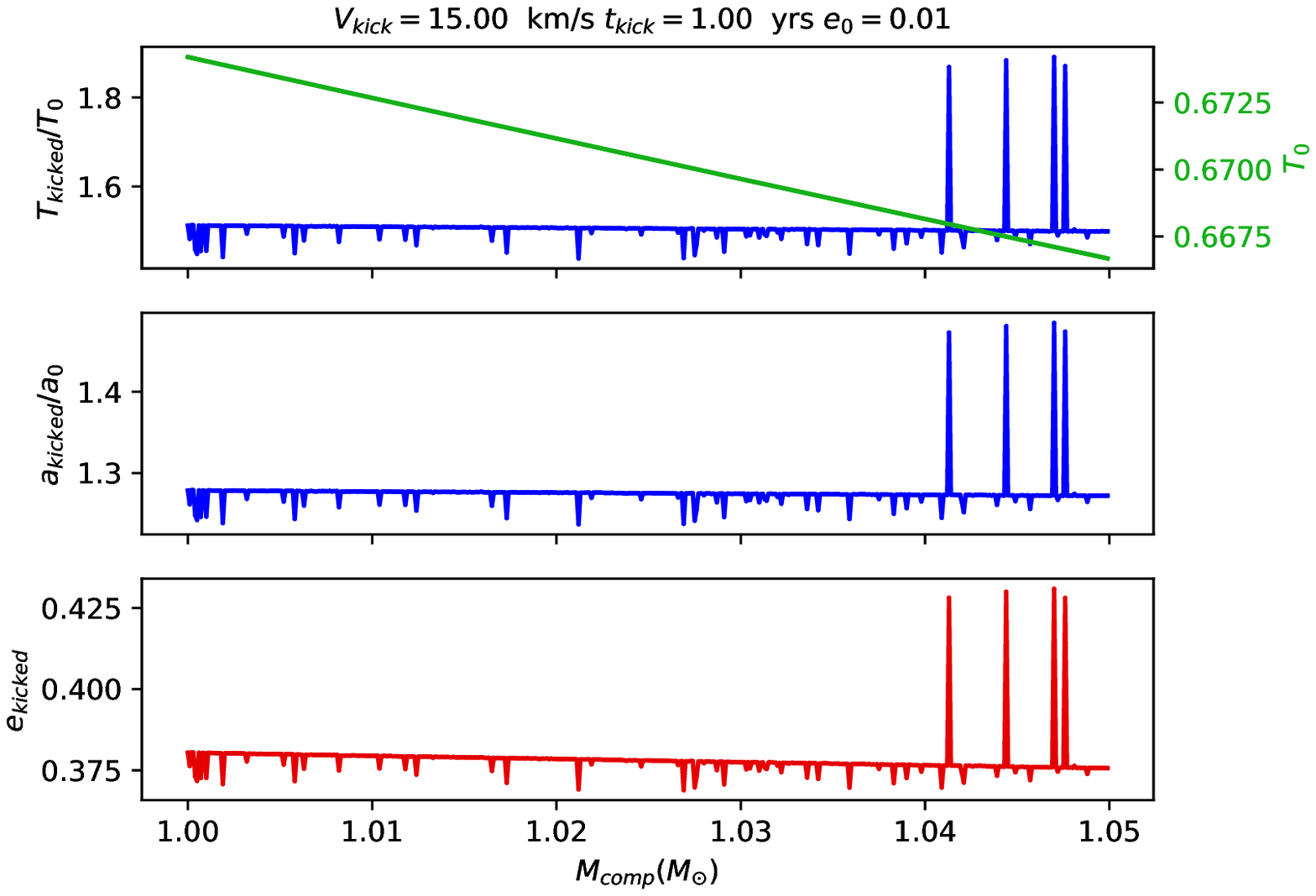} &
\includegraphics[width=0.5\linewidth]{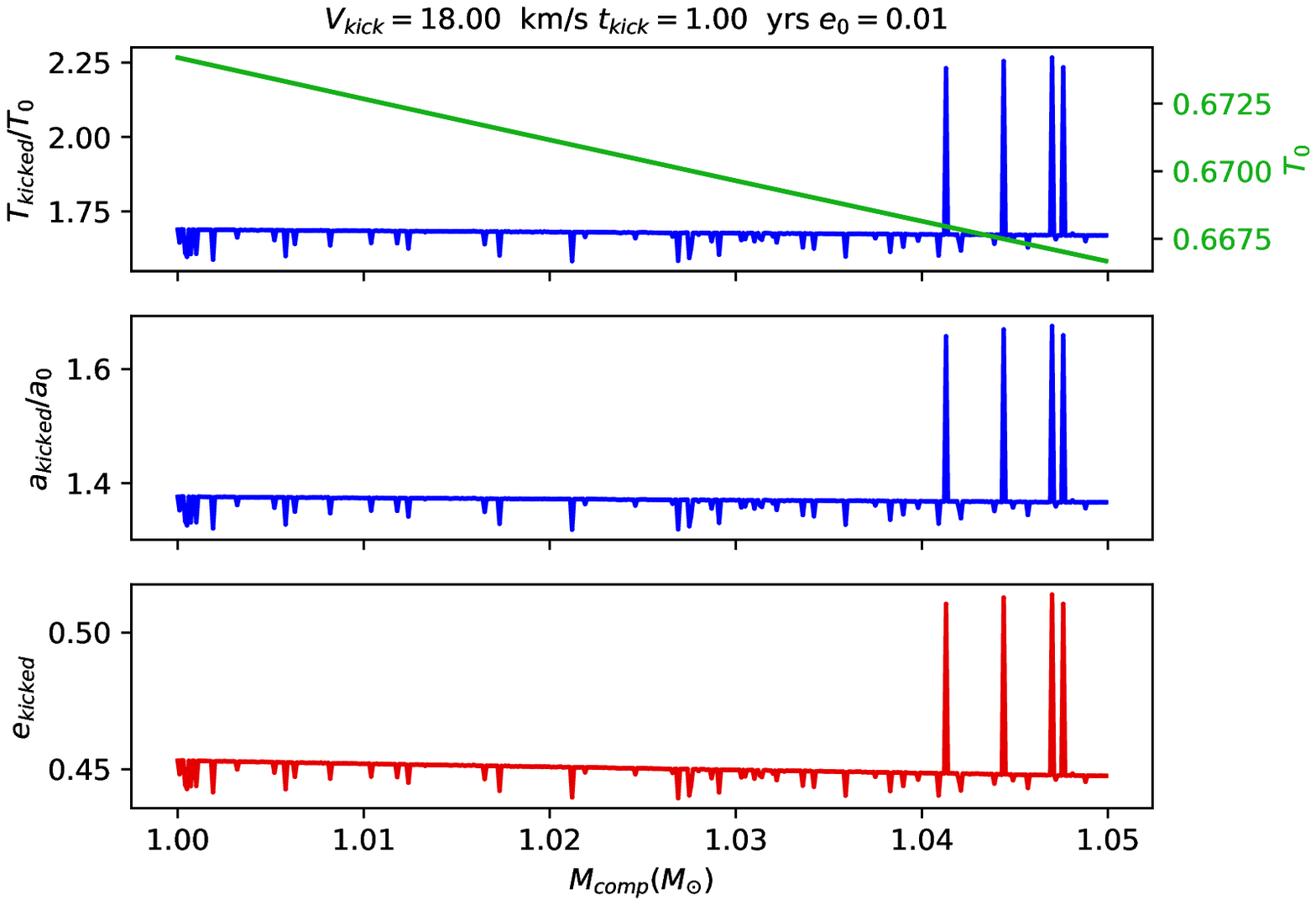}\\
\includegraphics[width=0.5\linewidth]{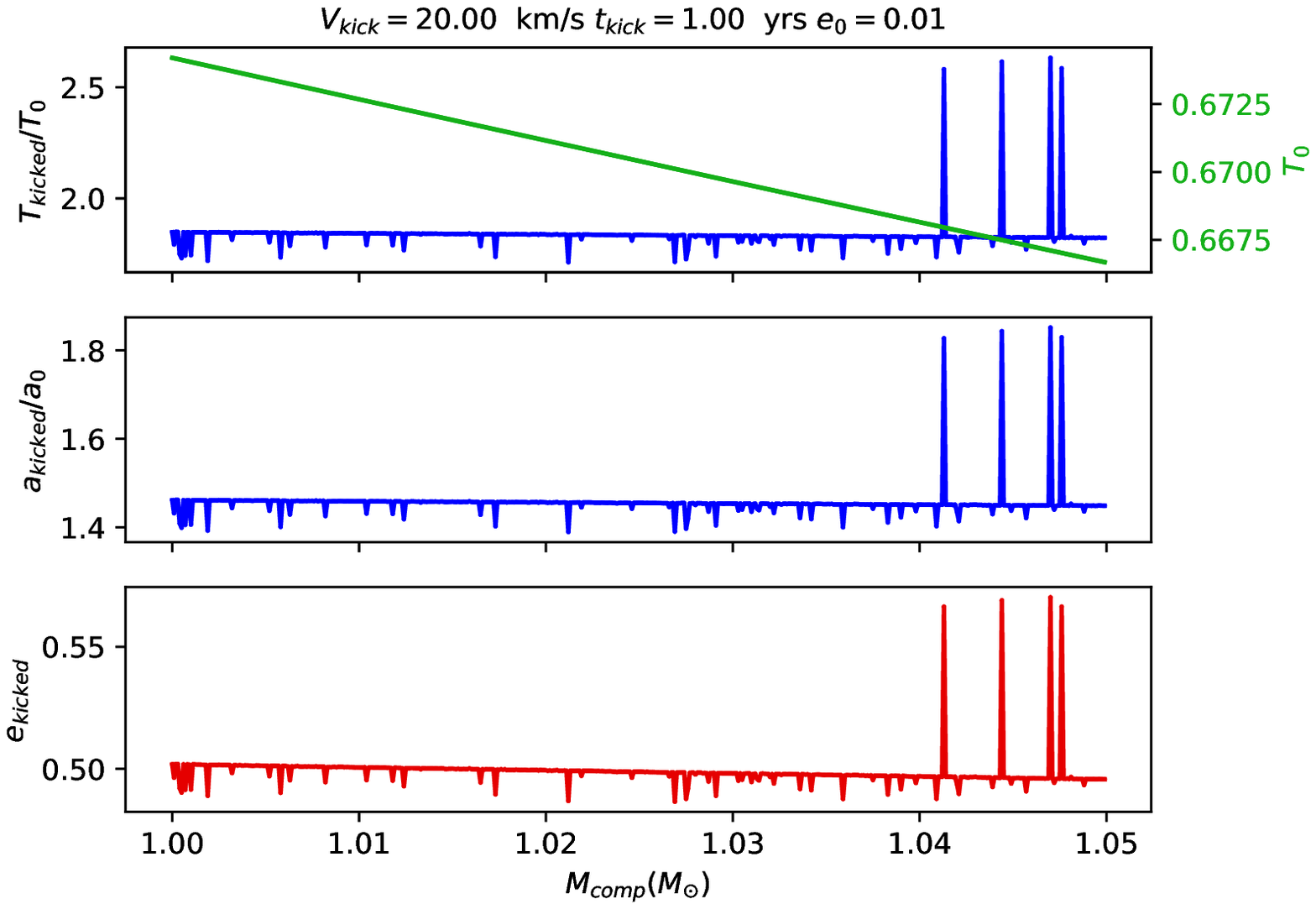} &
\includegraphics[width=0.5\linewidth]{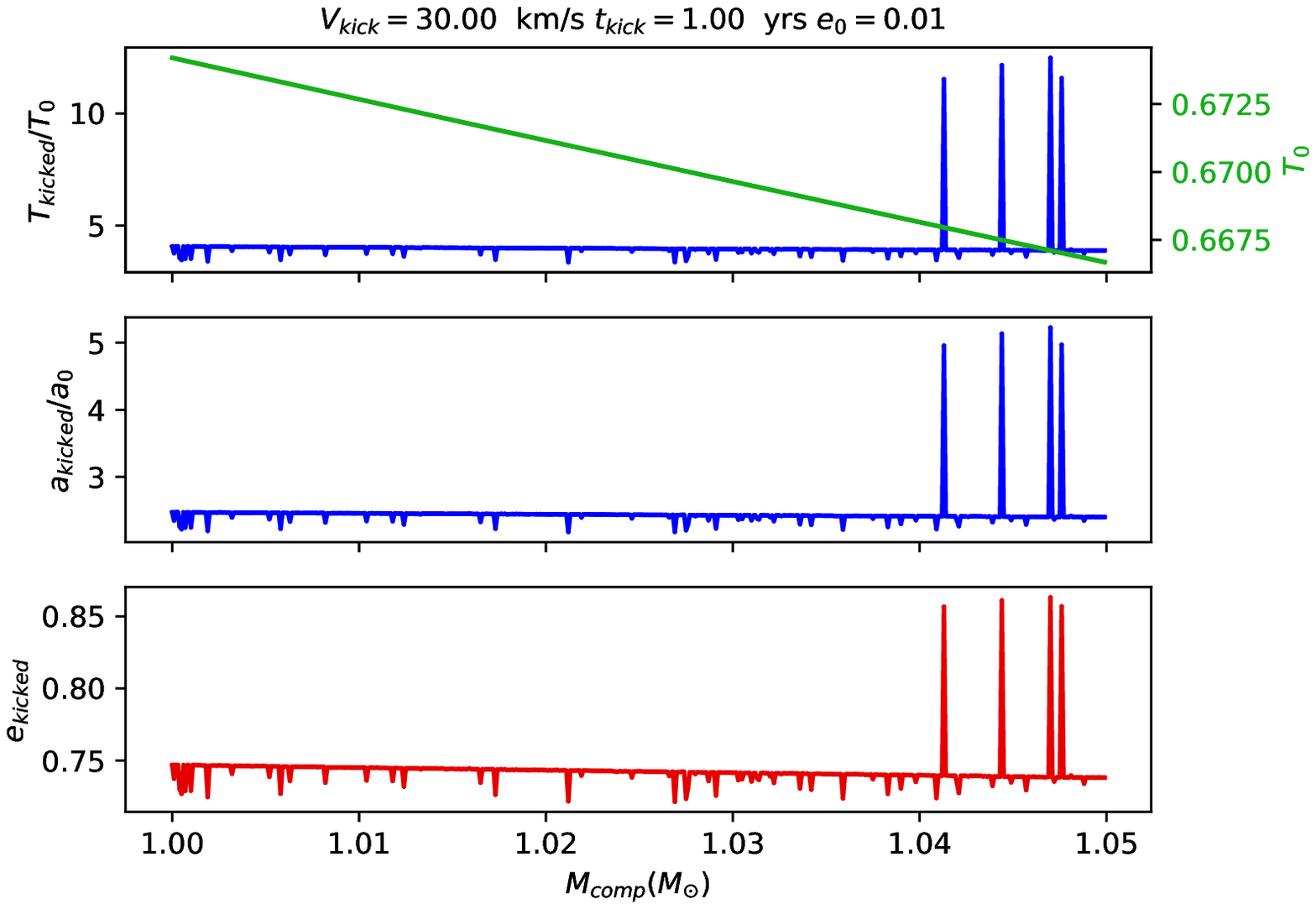}\\
\end{tabular}
\end{center}
\caption{The consequences of different kicks
on a typical binary system, as the initial
mass $M_{comp}$ slightly increases and remaining
parameters are kept constant.
The Figure contains four panels, corresponding to
four different $v_{kick}$ velocities.
Each panel has an upper diagram,
showing the resulting period $T_{kick}/T_0$
in blue. For reference,
the initial period $T_0$ is showed in green.
$T_0$ decreases as the $M_{comp}$ increases.
The middle diagrams
show the resulting semi-major axes $a_{kick}/a_0$
(in blue). Finally, the bottom diagrams, in red,
show the resulting values of the eccentricity
of the system after the kick.
One can see complex patterns,
resulting of the strong dependency
on initial conditions.}
\label{kicker_explore}
\end{figure*}

The Figure~5 analyses the
consequences of a kick as we vary the initial
mass $M_{comp}$. We will keep constant the remaining  parameters
aiming to explore the strong sensitivity
to initial conditions.
The Figure contains four panels, corresponding to
four different $v_{kick}$ velocities.
Each panel has an upper diagram,
showing the resulting period $T_{kick}/T_0$
in blue. The initial period $T_0$ in green,
for reference ($T_0$ decreases as the $M_1$ increases).
The middle diagrams
show the resulting semi-major axes $a_{kick}/a_0$
(in blue).
Finally, the bottom diagrams, in red,
show the resulting values of the eccentricity
of the system after the kick.
One can see complex patterns,
resulting of the strong dependency
on initial conditions.

As expected, the Figure~5 shows that
larger periods, longer semi-major axes and
larger eccentricity values are obtained as
$v_{kick}$ increases.
Indeed, beyond certain $v_{kick}$, the
system might be left un-bounded in some cases.

Interestingly,
the Figure~5 shows
a complex pattern of spikes.
These spikes reflect a variety of changes in the
final orbital parameters when the same explosion is
applied to the initial systems
defined by different $M_{comp}$.
In other words, the WD
explodes at the same instant in all cases. Therefore, the
kick occurs at the same time in all modeled binaries.

Figure~5  shows that even very small variations
in the control parameter $M_{comp}$ produces very different
results. This is because  as $M_{comp}$ changes,
the $v_{kick}$ is added to systems with slightly different orbital periods and different orbital velocities.
One gets different results as $M_{comp}$ varies, and the system shows
a strong dependency on initial conditions.

This pattern
indeed resembles a fractal-like structure.
The Figure~5 shows a variety of
interleaved small and large changes, and also interleaved
increments and decrements in the final values of the
semi-major axes and eccentricities.
In short, different $T_0$ values make
or not the whole system more sensitive to the
kick, even when the explosion occurs at the same time in all
cases. The stars are in different
relative positions at $t_{kick}$ and
one adds kick vectors pointing to different
directions when the kick takes place.
Hence, one can get a complex behavior resulting
from both dependencies.

%

\textbf{\section{Discussions}}

Our method based on the equations gives analytical expressions for parameters of the binaries after asymmetric kick
and probability of disruption of a binary
depending on the ratio of kick velocity and velocity of components prior
to explosion and on relative amount of the ejected matter. We assume here the explosion to be instantaneous (infinitely short duration) and would takes place in very close binaries, in the range of LMXBs and close double pulsars, such as the PSR 1913+16, J0737- 3039AB and the
close WD-pulsar system PSR 0655+64 (van den Heuvel 2009). As a result, $v_{final}\approx v_{initial}\approx0.1c$.
To analytically find the separation distance at the moment of explosion, we considered a system in which the initial conditions are
$\Delta M = 0.01M_{\odot}$, $M_{WD} = 1.1 M_{\odot}$ and $M_{comp,i} = 3 - 6 M_{\odot}$.

In Figure~2, we note that several values of $\eta$ appear to correlate with the orbital parameters.
In addition, our results cover the observational orbital parameters up to $P_{orb} \geq 90$ d.
We should emphasize here that the degree of interaction is critically dependent on the critical value of $\eta$. At $\eta < 0.40$
the kick velocity has very small effect on the orbital distributions, as we can show that M$_{ej}$ is on the order of 0.1 (the mass "lost" in the AIC itself). On the other hand, when $\eta \geq 0.40$ this would lead
to a significant increase of their orbital distribution. This ratio also serves to demonstrate the sensitivity of binary evolution,
and has also crucial characteristics in examining influence of physical parameters on the binary evolution.  It has recently been argued that,
the majority of the observed double NS systems can best be explained with small kicks in the second SNe (Tauris et al. 2017).


\begin{figure*}
\begin{center}
\begin{tabular}{cc}
\includegraphics[width=0.5\linewidth]{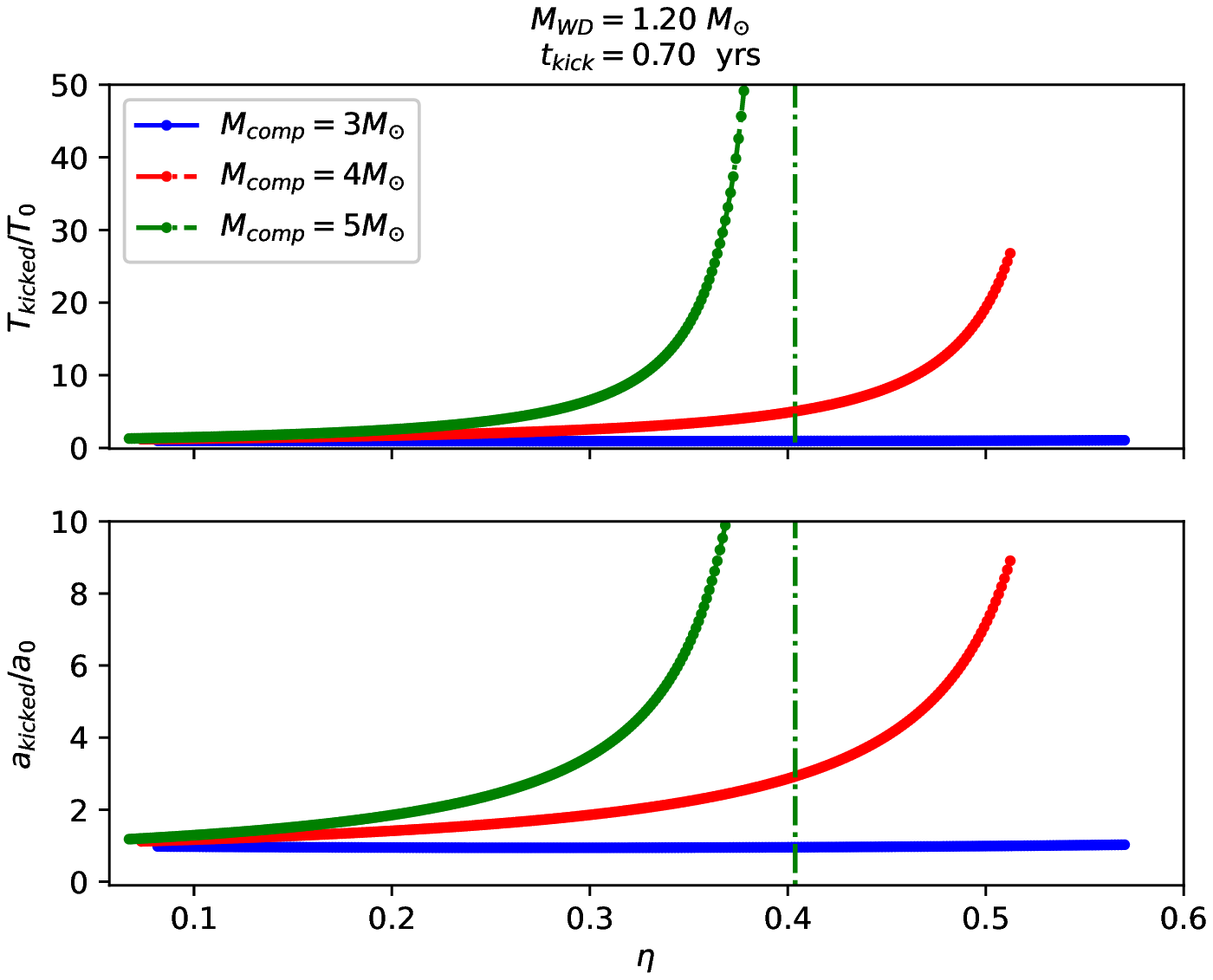} &
\includegraphics[width=0.5\linewidth]{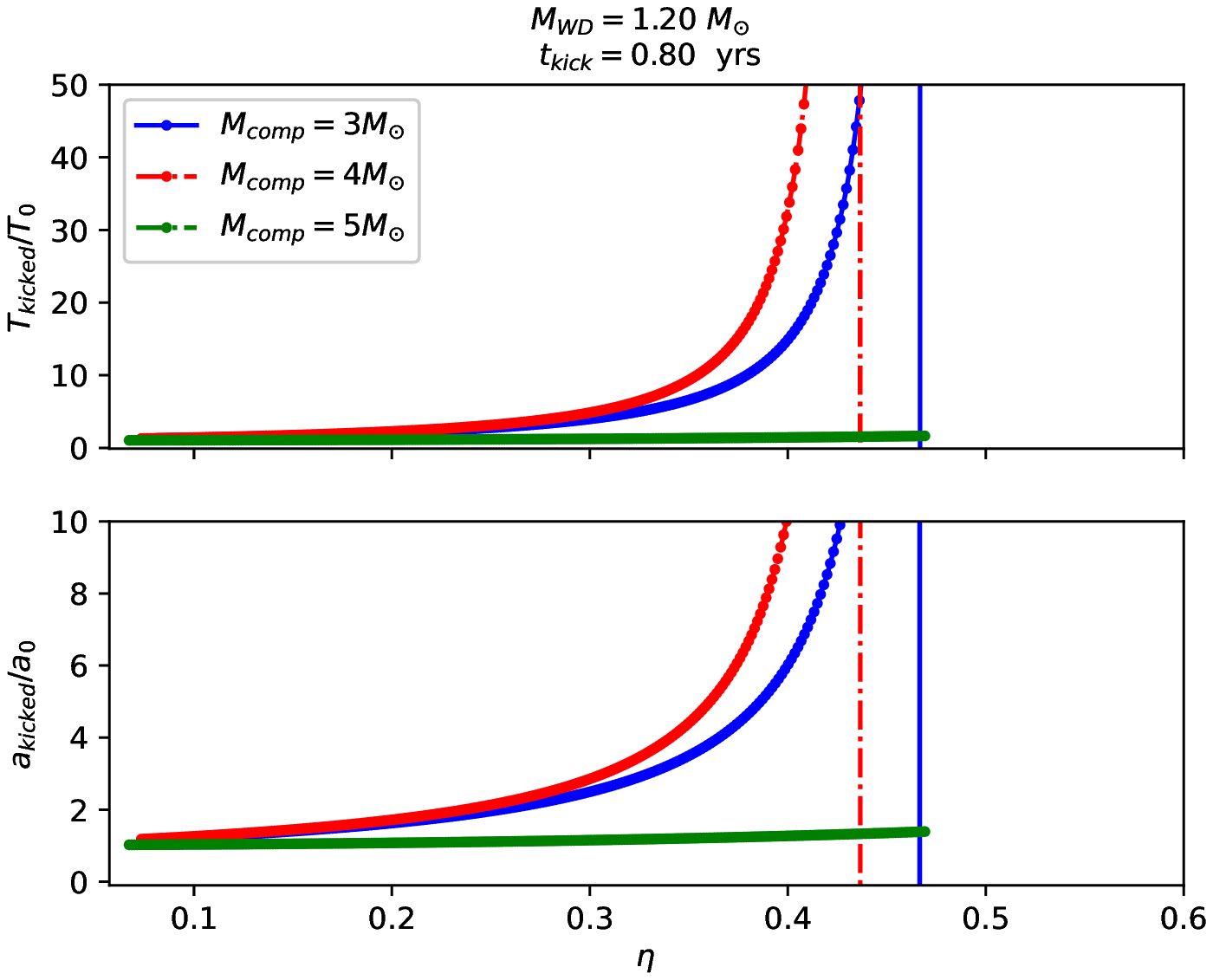}
\end{tabular}
\end{center}
\caption{The consequences of different kicks
on a binary system, with kicked star
fixed to be $M_{WD} = 1.2 M_{\odot}$,
and three different values of $M_{comp}$, hence, of $T_0$.
The leftmost diagram corresponds to $t_{kick}=0.7$ and
the rightmost diagram corresponds to  $t_{kick}=0.8$.
In each diagram, top panel show the evolution
of $T_{kick}/T_0$ with $\eta$, while
bottom panel show the evolution of $a_{kick}/a_0$.
}
\label{eta_explore}
\end{figure*}

The Figure~6 shows
the consequences of different kicks
on a binary system, with a kicked star
fixed to be $M_{WD} = 1.2 M_{\odot}$,
and three different values of $M_{comp}$, hence, of $T_0$.
The two diagrams correspond to $t_{kick}=0.7$ and $0.8$.
In each diagram, top panel show the evolution
of $T_{kick}/T_0$ with $\eta$, while
bottom panel show the evolution of $a_{kick}/a_0$.

The shapes of the curves are roughly similar,
following the same evolution that was described
in the analytical study.

In general, these curves show that the final period and semi-major axis increases with $\eta$ (that is, with $v_{kick}$). However, this increase
is stronger for certain values of $M_{comp}$.
Notably,
for each $M_{comp}$, there is a different $\eta$ limit value
which makes the system unbounded, reflected by the
vertical line.  
One can also see that the results are different for each kick time, as the leftmost and rightmost diagrams show different limit $\eta$ values.


Hence, the Figure~6 is reflecting again
the strong dependency to the initial control parameter values.
There is no simple evolution of the plotted curves
for each $M_{comp}$. One observes that
the green curve corresponding to
the largest mass $M_{comp}=5.0$ $M_{\odot}$ flattens
with the increase in $t_{kick}$ value, reflecting smaller changes
in the final orbital parameters. Conversely, the blue curve
corresponding to the smallest mass $M_{comp}=3.0$ $M_{\odot}$, 
rises as the $t_{kick}$ value increases. This reflects
larger changes in the final orbital parameter as the $\eta$ parameter
grows in these cases.

Because this complexity,
further numerical explorations on the parametric
space are needed as a natural extension to the work done.


Further observational evidence that at least some NSs receive low kicks at birth. The observed of pulsations with a period of 11.1 ms in the accreting
binary, J0900-3144 (Burgay et al. 2006), has
confirmed this scenario. Additionally, Champion et al. (2008) proposed that the AIC of a massive
and rapidly rotating WD ($M$ = $1.05_{\odot}\pm0.015_{\odot}$) could produce the observed orbital
parameters of PSR J1903+0327 ($M_{NS}$= $1.667_{\odot}\pm0.021_{\odot}$). Recently, Freire et al.
(2011) and Pijloo et al. (2012) have questioned the plausibility
of the existence of such a MSP-WD-MS triple system
and have pointed out several shortcomings associated with
such a proposition. Further challenges the conventional formation scenarios not only is the
pulsars companion a main-sequence star, but the orbit is also highly
eccentric (e = 0.44) with 95-d orbital period, neither of which are predicted by the standard
formation scenarios. On the hand, Abdikamalov et al. (2010) have presented results from an
extensive set of general-relativistic AIC simulations using a microphysical
finite-temperature equation of state during collapse; they investigated a set
of 114 progenitor models in axisymmetric rotational equilibrium, with a
wide range of rotational configurations and resulting WD masses. They contrast the gravitational-wave signals of AIC and rotating massive
star iron core collapse and find that they can be distinguished.
\\

\section{Conclusions}

We present a model to estimate the new orbital parameters relying
on the dynamical effects of the asymmetric mass ejection, during the accreting WD binaries in the initially tight binary system in the AIC process. Assuming the kick velocity axis coincident with the spin axis of the newborn NS, whose rotation is presumably co-aligned with the orbital angular momentum of the pre-explosion binary.
As a result, the instability here is associated with many dissipative processes such as kicks, mass loos and shock waves. These effects imparted on the newborn pulsars,  can account for the differences in their orbital
parameter distributions. Our results are valid for all orbital periods of MSPs up to $P_{orb,f} \leq 90~d$ in circular orbits in which the complete co-rotation
has achieved. While the final separation distance is more than three times as large as it's initial value.

Consequently, it is seen that all orbits retain their circularization characteristics after AIC process along the stellar
evolution during the instantaneous mass loss erasing any remnant eccentricity. The orbital period shows a steady increase in the magnitude, before the ratio of kick to escape ($\eta$) is reached a value of 0.40. After this stage, $\eta \geq 0.40$, it is anticipated that the distribution of the kick velocities rising to a sharp peak and would produce longer orbital periods. This ratio serves to demonstrate the sensitivity of binary evolution, since in all cases, the kick to the escape is smaller than the original orbital velocity in the fate of the binary. The magnitude and circularization of the orbits are determined by the nature of the companion star which we considered $3M_{\odot}\leq M_{com} \leq5M_{\odot}$ at the evolution of a binary under the instantaneous mass loss. However, the initial separations should be distributed according to the observed distribution of semi-major axes.

However, a small portion of mass ratio ($M_{WD}/M_{comp}$) could be affected to the
P$_{orb,f}$ and it exhibits considerable non-diversity in behavior. By comparing results from our model with observed binary MSPs, we can enhance our understanding of their formation, evolution and distributions.
Therefore their total contribution is expected to be around $20 \%$ (Taani et al. 2012a) of population, since many objects might have formed in such an environment. The distribution  functions  of  systemic velocities, correlation with orbital parameters and other effects associated with dynamical characterisation of MSP binary parameters can be studied in terms of constrain the Galactic potential and their modeling spatial distribution in the Galaxy (Kalogera 1996), and also  in  globular  clusters by assuming their smallest possible systemic velocity.


In addition, we have approached a more complete
scenario and numerically explored the case of planar, but
non-circular binaries. We have implemented
a very simple rotating mechanism to show the important role
played by the direction of the kick in the resulting
kicked system. Despite this simplicity, we have seen that the
combination of a variable kick time and a variable
kick vector direction leads to a complex behavior,
as different kick velocity components are combined with
different velocity components of the kicked star.
We have observed that even
for  initial small eccentricities and semi-major axes, 
the analysed problem seems to present strong sensitivity
to initial conditions. This means to observe a
complex behavior, with very different results for every set
of control parameters.

This strong sensitivity lead to a complex behavior,
reflected in the plots showing the dependency of the
new orbital parameters with the mass of one of the stars.
Indeed, these plots resemble fractal-like behavior.
The obtained patterns
are similar to those obtained in some other scattering
problems with presence of chaos, mainly plots of
differential cross sections or delay-time functions (Seoane et al. 2013).  The presence of this complex behavior in a
deterministic system is a sign of strong sensitivity to initial conditions, but it is not a synonym of fully chaotic motion.
Assessing the presence of chaos is a necessary step for classifying the resulting orbits in regular orbits
or chaotic ones.
Hence, further work is still needed to formally characterise this complexity. This is an interesting research to extend the results of our work.


Although, the prevalence of AIC is still a matter of heated debate both observationally and theoretically. However, our results are promising approach, and more stringent constraints can be applied to the kick velocity and mass-loss and separation between the components and the new formed binary. This will lead  understanding the binary stellar evolution with the birth of the binary and double pulsar systems. More observations are required to explain these features and behaviors.

\section*{Acknowledgements}

A. Taani would like to thank NAOC and IHEP institutes for supporting his visit as Visiting Scientist Researcher.
Juan C. Vallejo thanks the Spanish Ministry of Economy, Industry and Competitiveness for grants ESP2017-87813-R.
We are grateful to Alexander Tutukov, Postnov Konstantin, Manjari Bagchi, Iminhaji Ablimit and Serena Repetto for their
comments and suggestions that allowed to improve the clarity of the
original version.
The authors would also like to thank the anonymous referee
for the careful reading of the manuscript and for all suggestions and comments which
allowed us to improve both the quality and the clarity of the paper.


\section*{DATA AVAILABILITY}
The data that support the findings of this study are available
from the corresponding author upon reasonable request.

\end{document}